\documentclass[twocolumn,showpacs,preprintnumbers,amsmath,amssymb]{revtex4}

\usepackage{graphicx}
\usepackage{dcolumn}
\usepackage{bm}

\begin{document}


\title{Sub-Cycle Strong-Field Interferometry}
\author{Christian~Ott}
\author{Philipp~Raith}
\author{Thomas~Pfeifer}
\email{tpfeifer@mpi-hd.mpg.de}
\affiliation{Max-Planck-Institut f\"ur Kernphysik, Saupfercheckweg 1, 69117 Heidelberg, Germany}

\date{\today}

\begin{abstract}
A nonlinear interferometry scheme is described theoretically to induce and resolve electron wavefunction beating on time scales shorter than the optical cycle of the time-delayed pump and probe pulses.  By employing two moderately intense few-cycle laser fields with a stable carrier-envelope phase, a large range of the entire electronic level structure of a quantum system can be retrieved.  In contrast to single-photon excitation schemes, the retrieved electronic states include levels that are both dipole- and non-dipole-accessible from the ground electronic state.  The results show that strong-field interferometry can reveal both high-resolution and broad-band spectral information at the same time with important consequences for quantum-beat spectroscopy on attosecond or even shorter time scales.\end{abstract}


                             
\maketitle
Measurement and control of electron quantum wavefunctions is a key goal of attosecond science, and of quantum physics in general.  Typically, the shortest flashes of coherent radiation (attosecond pulses)~\cite{GOULIELMAKIS2008} with photon energies of several tens of eV are employed to gain access to this electronic realm of dynamics.  In measurements, two attosecond pulses~\cite{SEKIKAWA2004} or an attosecond pulse in combination with a femtosecond pulse at a variable relative time-delay~\cite{CORKUM2007,MAURITSSON2010} are typically used to track ultrafast dynamics with attosecond resolution.

Interferometric methods have been employed for a long time to access the vibrational and rotational level structure in molecules~\cite{KHUNDKAR1990,SCHERER1991,FEUERSTEIN2007,WINTER2009}.  Experiments also revealed interferences of high-lying electronic Rydberg states in atoms~\cite{JONES1993} and, later, between free-electron wavepackets~\cite{WOLLENHAUPT2002}. A theoretical study recently showed that even for two-electron systems and single-photon excitation one can observe quantum interferences of free electrons~\cite{PALACIOS2009}.  A very recent experiment~\cite{MAURITSSON2010} extended the interferometric toolkit by employing isolated attosecond pulses for measuring bound and continuum electronic wavepackets.  In all these earlier interferometric approaches, the minimum resolvable wavepacket beating period was larger than the optical cycle of at least one of the pulses used or exactly a small integer multiple of that optical cycle.

In this letter, we present a fundamentally new route to inducing and probing electronic quantum-state interferences over a large spectral range on time scales shorter than the cycle of the light used for excitation and probing.  We describe, analytically and computationally, a comprehensive and general method to extract continuum- and bound-state electronic wavefunction beating in atoms on sub-femtosecond time scales, which does not require attosecond pulses nor ultraviolet or shorter-wavelength fields.  The method is based solely on the highly-nonlinear yet coherent strong-field excitation of electronic superposition states by few-cycle femtosecond laser pulses and the quantum-interferometric principle.  Strong-field interferometry is enabled by the advent of carrier-envelope phase (CEP) stabilized laser pulses that provide a temporally well-defined (to a small fraction of a femtosecond) electric field for the excitation and pump step.  We also show that strong-field nonlinear excitation and probing can be used to obtain temporal and spectral signatures of the field-free evolution of electronic wavepackets and thus a large portion of the entire electronic-level structure at high spectral resolution, without using multi-eV broadband light, nor relying on a widely tunable light source of high spectral purity.

To convey the key idea of strong-field interferometry, we start out by describing the wavefunction of an atomic system after its interaction with an intense femtosecond pump pulse.  Recent experiments showed that strong-field interaction of atoms with light efficiently populates excited electronic states~\cite{NUBBEMEYER2008,EICHMANN2009}.  For one-electron excitations (i.e. intensities slightly above the onset of strong-field single ionization), the electronic wavefunction of the pump-pulse excited atom can thus be written, in most general terms:
\begin{eqnarray}
\label{wavepacket}
|\Psi(t)\rangle=\sum_n b_n e^{-i E_n t} |\psi_n\rangle + \int_{\vec{k}} d^3k~c(\vec{k}) e^{-\frac{i}{2} k^2 t} |\vec{k}\rangle\nonumber,
\end{eqnarray}
where $E_n$ are the bound state energies, $b_n$ and $c(\vec{k})$ are the complex wavefunction expansion coefficients for the bound-state $|\psi_n\rangle$ and continuum-state wavefunctions $|\vec{k}\rangle$, respectively.
This superposition state evolves freely for a time $\tau$ (atomic units $\hbar=e=m_e=1$ are used throughout), given by the delay between the pump and the probe pulses.  After the probe laser pulse, the evolved initial superposition state is modified and its continuum portion can be written as
\begin{eqnarray}
|\tilde\Psi_c(t)\rangle&=&\int_{\vec{k}}d^3k~|\vec{k}\rangle\langle\vec{k}| \hat{L}_\mathrm{pr}|\Psi(\tau)\rangle e^{-\frac{i}{2} k^2 (t-\tau)}\nonumber\\
&=&\int_{\vec{k}}d^3k~ \tilde{c}(\vec{k})e^{-\frac{i}{2} k^2 (t-\tau)} |\vec{k}\rangle
\end{eqnarray}
where $\hat{L}_\mathrm{pr}$ describes the full coupling of the atom with the intense probe laser pulse $\tilde{E}_\mathrm{pr}(t)$. The $\tilde{c}(\vec{k})$ represent the amplitudes of the continuum wavefunction after the probe laser pulse:
\begin{equation}
\tilde{c}(\vec{k})=\sum_n c_n'(\vec{k}) e^{-i E_n \tau} + c''(\vec{k}) e^{-\frac{i}{2} k^2 \tau}
\end{equation}
with
\begin{eqnarray}
c'_n(\vec{k})&=&L_\mathrm{pr}^{\vec{k}n}b_n\nonumber\\
c''(\vec{k})&=&c(\vec{k})e^{-\frac{i}{2}\int_{t_1}^{t_2}\vec{A}(t)^2 dt},\nonumber
\end{eqnarray}
in which $L_\mathrm{pr}^{\vec{k}n}=\langle\vec{k}| \hat{L}_\mathrm{pr}|\psi_n\rangle$ stands for the matrix element of $\hat{L}_\mathrm{pr}$ connecting bound state $n$ with the momentum state $\vec{k}$ due to the probe pulse.  Times $t_1$ and $t_2$ are chosen such that the probe laser pulse is fully contained in the interval $[t_1,t_2]$ before and after the pulse, respectively, and the vector potential $\vec{A}(t)$ is defined in the Coulomb gauge with $\lim_{t\to\pm\infty} |\vec{A}(t)|=0$.  The photoelectron momentum spectrum thus reads:
\begin{eqnarray}\label{PhotoElectronSpectrum}
S(\vec{k})&=&|\tilde{c}(\vec{k})|^2\propto\sum_n |c_n'(\vec{k})|^2 + |c''(\vec{k})|^2  \\
&+&\left[\sum_{n,m<n} c_n'(\vec{k})c_m'^*(\vec{k}) e^{-i (E_n-E_m) \tau} \right. \nonumber \\
&+&\left. \sum_n c_n'(\vec{k}) c''^*(\vec{k}) e^{-i (E_n -\frac{1}{2} k^2) \tau} + c.c. \right].\nonumber
\end{eqnarray}
The terms in brackets are responsible for $\tau$-dependent oscillations of the photoemission probability with frequencies $E_n-E_m$ and $E_n -\frac{1}{2} k^2$.  Note that these interference terms appear \emph{regardless} of the nature or complexity of the pump and probe interaction.  For few-cycle pulse excitation as considered below, the CEP enters the complex-valued wavefunction coefficients $c_n'(\vec{k})$ and $c''(\vec{k})$ such that for non-CEP stable pulses the interference patterns are generally washed out.

The physical picture of the process is the following:  A first intense laser pulse (pump) excites a coherent superposition of electronic states (both bound and continuum) before the second intense pulse (probe), arriving at time delay $\tau$, projects these states onto common final continuum states that can be detected using photoelectron spectroscopy.  The coherently excited population of electronic states created by the pump pulse evolves with different phases (due to the different energies of the populated states) and thus leads to an energy-specific interference pattern in the final-state amplitudes as a function of pump--probe delay time as indicated by Eqn.~(\ref{PhotoElectronSpectrum}).  Even though both the pump and the probe step are nonlinear and can be complex, the evolution of the states proceeds in the field-free temporal window between the two pulses, resulting in an interference pattern that is governed by the field-free evolution of the quantum system.  Note that the strong-field nature of the pump step does not limit us to transitions among odd-parity electronic states (as for a single-photon pump step~\cite{MAURITSSON2010}) but, in principle, all energetically allowed states can be accessed, e.g. by means of multiphoton or (frustrated) tunneling processes~\cite{NUBBEMEYER2008}.

\begin{figure}
\centering
\includegraphics[width=\linewidth]{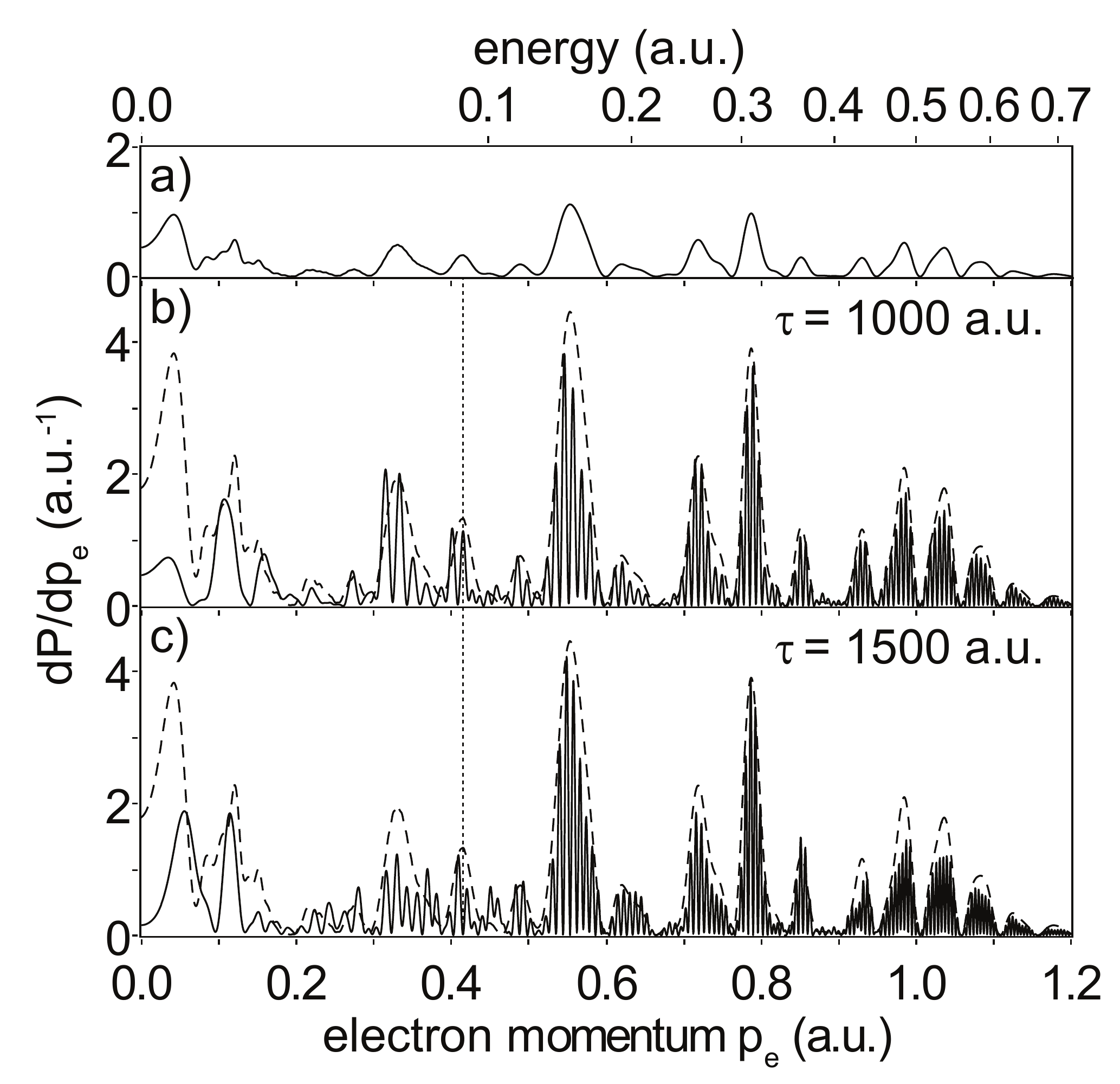}
\caption{\label{Figure1_ATI_Spectra}
Photoelectron momentum spectra a) for one single pulse, b) and c) for two identical pulses at two different time delays $\tau$ (solid lines). In b) and c), the rescaled (times 4) single-pulse spectrum of a) is also shown (dashed line).  A vertical dotted line at an electron momentum $p_e=0.415$ is drawn to highlight the time-delay-dependent modulation.}
\end{figure}

In the following, we describe a numerical simulation that was carried out to computationally support the above theory.  An atom with one single active electron is considered here, using the soft-core binding potential $V(x)=-(x^2+a^2)^{-1/2}$ introduced by Javanainen et al.~\cite{JAVANAINEN1988} which has since then become a routine approach to study the interaction of atoms with strong laser fields.  The parameter $a$ is chosen to match the ground state energy $E_0=-15.7596~\textrm{eV}$ ($a\approx1.154$) of argon. The non-relativistic, dipole-approximation Hamiltonian with the laser field present thus reads $\hat{H} =(\hat{p}+A(t))^2/2-1/\sqrt{\hat{x}^2+a^2}$ with the vector potential $A(t)$ defined in the Coulomb gauge. The split-operator technique~\cite{FLECK1976,FEIT1982,PFEIFER2004A} is employed for the numerical evolution of the time-dependent one-dimensional Schr\"odinger equation. The ground state of the system is obtained by field-free evolution in imaginary time ($\Delta t\rightarrow -i\Delta t$).

Laser pulses with a duration of 6~fs, a carrier wavelength of 800~nm and a CEP corresponding to a sine-like field are used in the simulation.  The  peak of the electric field envelope is $\tilde{E}_{max}=0.08$, corresponding to a peak intensity of $2.2\times10^{14}~\textrm{W}/\textrm{cm}^2$.  Two equal pulses of this kind are temporally delayed with respect to each other for a range of $\tau$ between 1000 and 2998 atomic units of time.  For each delay, the wavefunction is evolved for a time of 4000~au with a step size of 0.1. The parameters of the space grid ($N=512,~\Delta x=0.5$) are chosen such that recolliding electron trajectories are fully contained.  An imaginary absorbing potential is used to collect the outgoing electrons in order to obtain the photoelectron spectra.

\begin{figure}
\centering
\includegraphics[width=\linewidth]{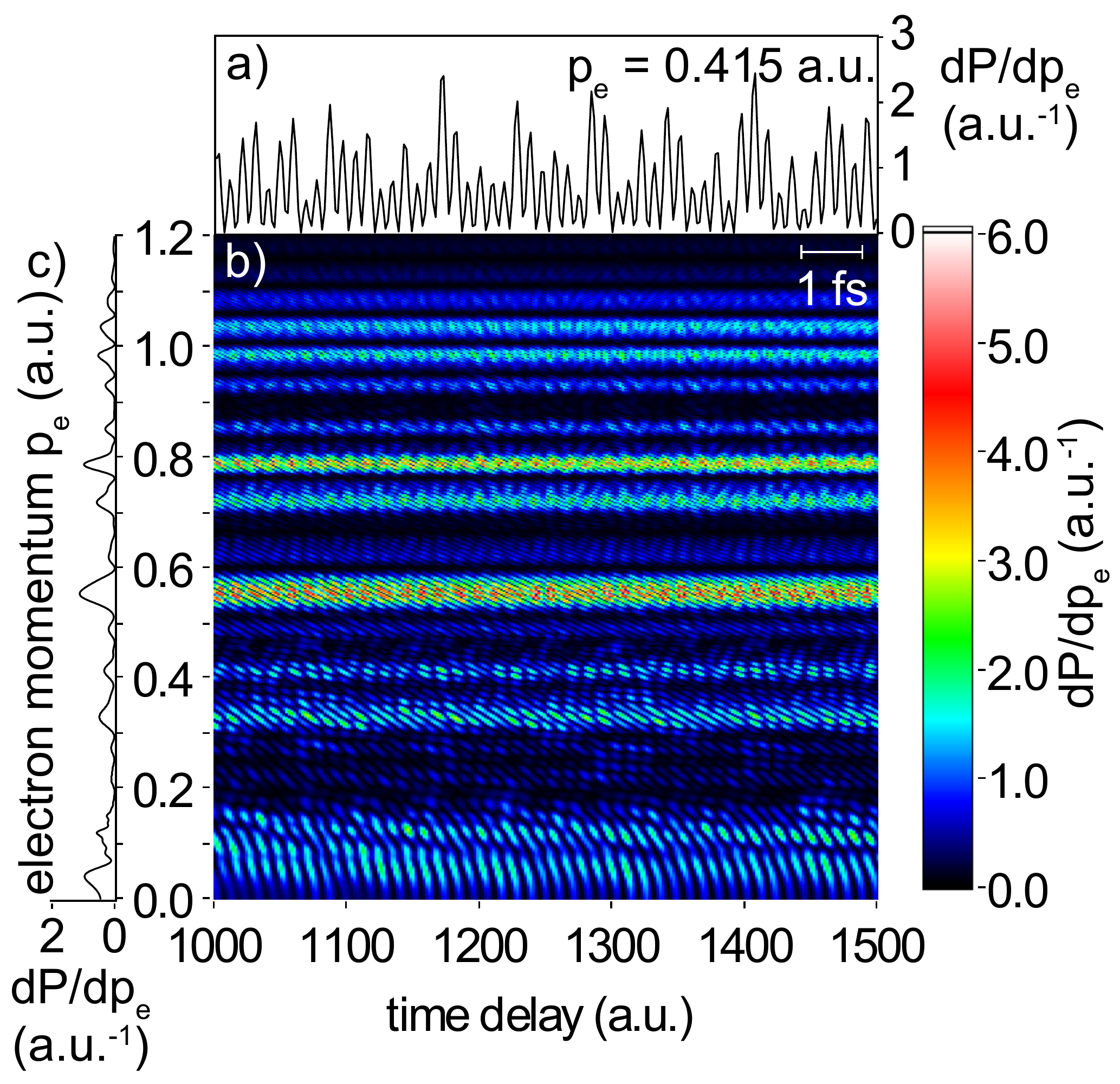}
\caption{\label{Figure2_TimeDelaySpectra}(color online)
Delay-time dependent photoelectron spectrum a) at the selected photoelectron momentum $p_e=0.415$ (dotted line in Fig.~\ref{Figure1_ATI_Spectra}), b) for a broad range of photoelectron momenta. The single-pulse spectrum from Fig.~\ref{Figure1_ATI_Spectra}a is additionally shown in c).  Attosecond time-scale beatings are clearly observable as a function of time delay.  Note the momentum-dependent visibility of the modulation in (b) corresponding to the peaks and dips of the single-pulse photoelectron spectrum c).}
\end{figure}

In Fig.~\ref{Figure1_ATI_Spectra} the obtained photoelectron momentum spectra are plotted both for a single pump pulse as well as for two pump--probe cases at different time delays $\tau$. Spectral modulations that are enveloped by the single pulse spectrum are observed. This modulation is time-delay dependent as predicted by Eqn.~(\ref{PhotoElectronSpectrum}). The entire set of photoelectron spectra between $\tau_1$ and $\tau_2$ are shown in Fig.~\ref{Figure2_TimeDelaySpectra}.
Even though both pump and probe pulses are significantly longer than 1~fs, as is the optical cycle of 2.7~fs, pronounced non-trivially structured sub-femtosecond beating patterns are observable as a function of time delay.  This sub-cycle beating is due to the time-dependent complex electronic wavepacket created by and synchronized to the CEP-stable strong electric \emph{field} of the pump pulse that gets projected into the continuum by the probe field.   Importantly, this is in contrast to the temporally less well-defined excitation and probing by the longer pulse \emph{envelope} as for single-photon excitation or schemes using non-CEP-stabilized strong-field excitation that tend to wash out such sub-cycle beating patterns.

\begin{figure}
\centering
\includegraphics[width=\linewidth]{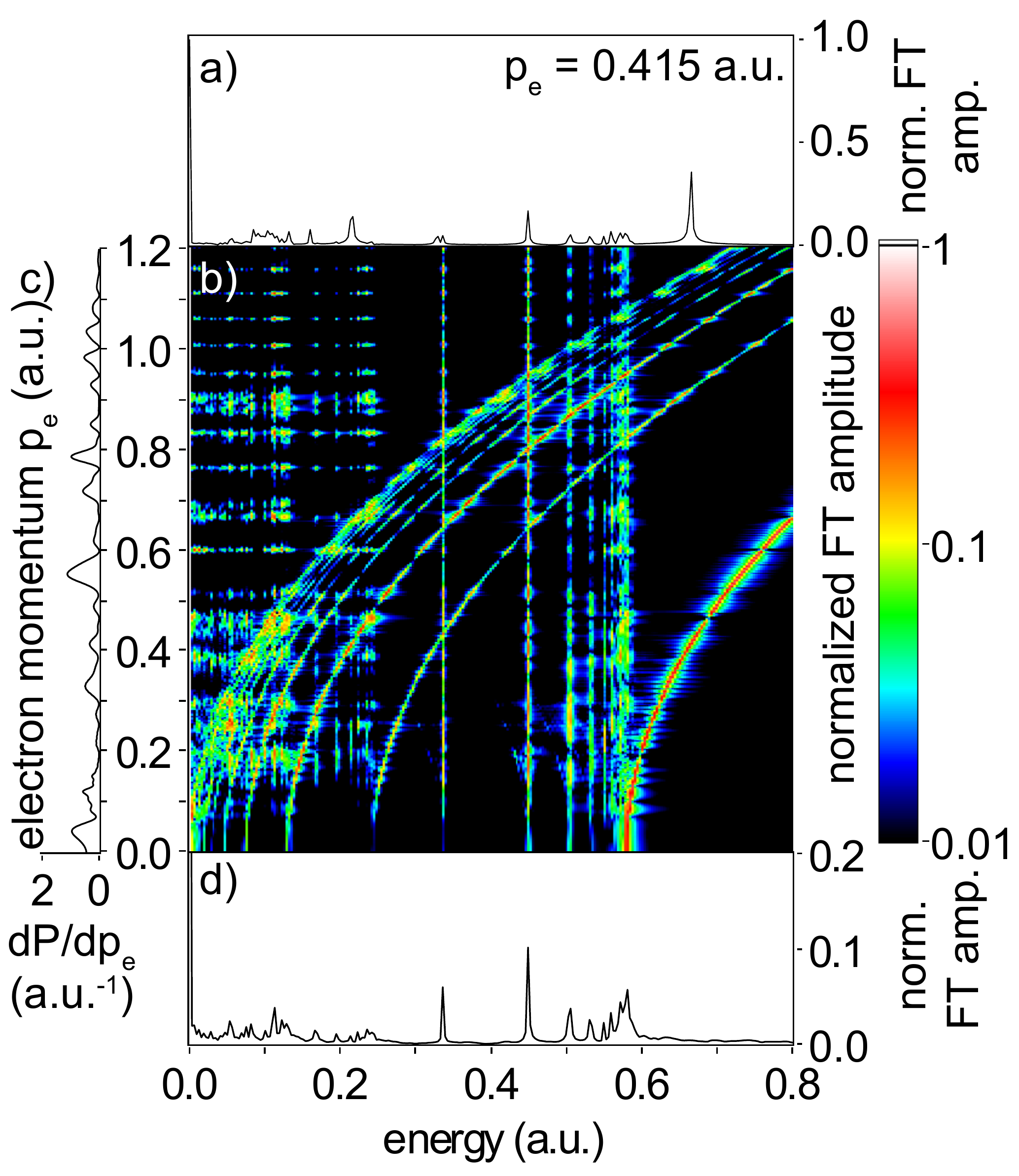}
\caption{\label{Figure3_LogFFTSpectra}(color online)
Beating frequencies of electronic bound and continuum states after Fourier transform of the data in Fig.~\ref{Figure2_TimeDelaySpectra}b: a) Lineout at momentum $p_e=0.415$, b) for all photoelectron momenta. Each Fourier-transformed spectrum is normalized to its zero-frequency component. c) The single-pulse photoelectron spectrum from Fig.~\ref{Figure1_ATI_Spectra}a.  d) Momentum-integrated frequency spectrum (see text for details).}
\end{figure}

\begin{figure}
\centering
\includegraphics[width=\linewidth]{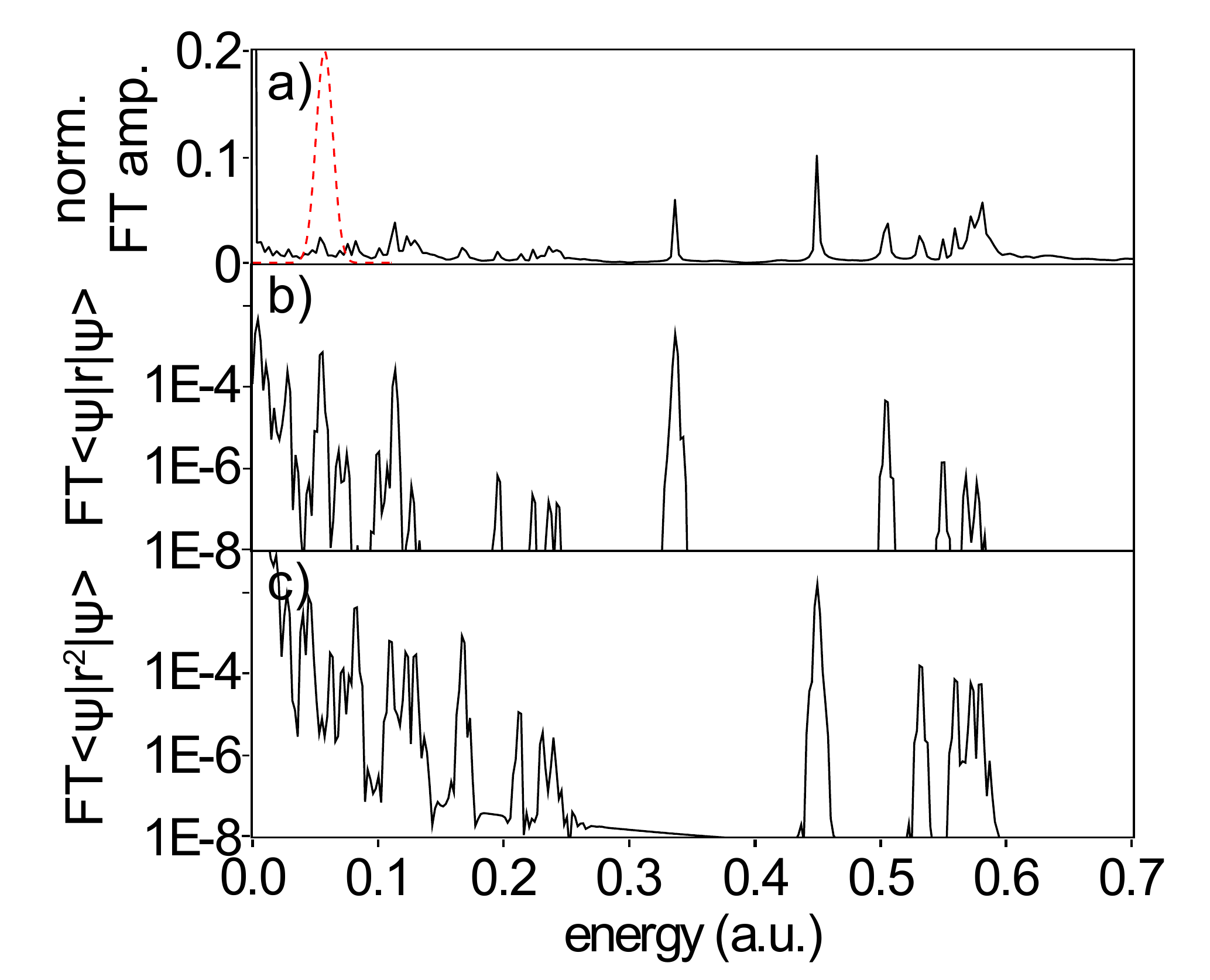}
\caption{\label{Figure4_DipoleQuadrupole}
(color online) a) Momentum-integrated frequency spectrum of the electron beating signal (from Fig.~\ref{Figure3_LogFFTSpectra}), (dashed red line: laser pulse spectrum).  Spectrum of the electron position (b) and position-squared (c) expectation value of the electronic wavefunction after interaction with the pump pulse, corresponding to the dipole- and quadrupole-active transitions, respectively.  The agreement of the spectral positions of the wavepacket beatings in b) and c) with the observable data from a) is excellent.}
\end{figure}

To reveal the beating frequencies within the time-delay modulations in Fig.~\ref{Figure2_TimeDelaySpectra}b, the photoelectron momentum spectra are Fourier transformed with respect to the time delay $\tau$. The result is shown in Fig.~\ref{Figure3_LogFFTSpectra}, exhibiting the entire set of beating frequencies predicted by Eqn.~(\ref{PhotoElectronSpectrum}). The modulation $|E_n-\frac{1}{2}k^2|$ (with negative binding energies $E_n$) is given by the parabola-shaped features which start at the corresponding energy level $|E_n|$ at zero photoelectron momentum. The strongest contribution to the modulation is observed in the parabola starting at the ground-state ionization potential energy $I_p=0.58$. This corresponds to an interference of free electron wavepackets ionized by the pump pulse and the probe pulse out of the ground state, as first measured by Wollenhaupt~\textit{et~al.}~\cite{WOLLENHAUPT2002}. The momentum-independent vertical lines can be assigned to $E_n-E_m$ and therefore refer to the attosecond beating between different bound electronic states of the atom.  A similar quantum beating among excited states has recently been observed experimentally for the case of a single-photon excitation scheme using an attosecond pulse~\cite{MAURITSSON2010}.  In our situation, for strong driving fields slightly above the onset of single-electron ionization, the ground state remains the most populated state in the system ($\approx93\%$ of the bound electron wavepacket) after the interaction with the laser pulses.  Thus the strongest vertical lines refer to electron wavefunction beating between the ground state and the higher excited states. Wavefunction beating among the higher excited states results in the weak vertical lines at low energies up to $E=0.24$ (the energy for ionization out of the first excited state).

To understand the nature of the electronic wavepackets created in the pump step, we examine the time-dependent expectation value of the position $\langle\Psi(t)|\hat{r}|\Psi(t)\rangle$ and the position squared $\langle\Psi(t)|\hat{r}^2|\Psi(t)\rangle$, i.e. the dipole-active and the quadrupole-active beating of the electronic wavepacket. Fig.~\ref{Figure4_DipoleQuadrupole} shows these dipole and quadrupole beating spectra, together with the momentum-integrated pump--probe spectrum from Fig.~\ref{Figure3_LogFFTSpectra}d. The position of the spectral lines, obtained from the pump--probe data (Fig.~\ref{Figure4_DipoleQuadrupole}a) agrees perfectly with the electron wavefunction beating (Fig.~\ref{Figure4_DipoleQuadrupole}b,c), proving that both dipole-allowed and dipole-forbidden transition energies among states are accessible within this approach.

The spectral resolution of the method is limited solely by the inverse of the pump--probe scanning period as in earlier interferometry approaches~\cite{SCHERER1991,JONES1993}.  In the example considered here, this corresponds to a spectral resolution of 40 meV, which is much less than the bandwidth of the individual femtosecond pulses (0.3~eV) and certainly much less than the entire spectral range of accessible beating frequencies ($>$16 eV).  This range is much larger than the photon energy of 1.6~eV---the spectral consequence of the sub-cycle temporal resolution of the strong-field interferometry scheme---creating new possibilities for quantum-beat spectroscopy on attosecond or even shorter time scales, independently on the availability of pulses of shorter duration.

Future applications of this technique can also include adding a third pulse temporally between the pump and probe pulses. Performing the analysis shown above, one can then extract, at high accuracy, even very small dressed-state energy shifts by observing phase shifts on the photoelectron beating signals.

We thank Thorsten Ergler for a helpful discussion and Robert Moshammer for a critical reading of the manuscript.  Financial support from the Max-Planck Research Group program is gratefully acknowledged.


\begin{thebibliography}{17}
\expandafter\ifx\csname natexlab\endcsname\relax\def\natexlab#1{#1}\fi
\expandafter\ifx\csname bibnamefont\endcsname\relax
  \def\bibnamefont#1{#1}\fi
\expandafter\ifx\csname bibfnamefont\endcsname\relax
  \def\bibfnamefont#1{#1}\fi
\expandafter\ifx\csname citenamefont\endcsname\relax
  \def\citenamefont#1{#1}\fi
\expandafter\ifx\csname url\endcsname\relax
  \def\url#1{\texttt{#1}}\fi
\expandafter\ifx\csname urlprefix\endcsname\relax\def\urlprefix{URL }\fi
\providecommand{\bibinfo}[2]{#2}
\providecommand{\eprint}[2][]{\url{#2}}

\bibitem[{\citenamefont{Goulielmakis et~al.}(2008)\citenamefont{Goulielmakis,
  Schultze, Hofstetter, Yakovlev, Gagnon, Uiberacker, Aquila, Gullikson,
  Attwood, Kienberger et~al.}}]{GOULIELMAKIS2008}
\bibinfo{author}{\bibfnamefont{E.}~\bibnamefont{Goulielmakis}},
  \bibinfo{author}{\bibfnamefont{M.}~\bibnamefont{Schultze}},
  \bibinfo{author}{\bibfnamefont{M.}~\bibnamefont{Hofstetter}},
  \bibinfo{author}{\bibfnamefont{V.~S.} \bibnamefont{Yakovlev}},
  \bibinfo{author}{\bibfnamefont{J.}~\bibnamefont{Gagnon}},
  \bibinfo{author}{\bibfnamefont{M.}~\bibnamefont{Uiberacker}},
  \bibinfo{author}{\bibfnamefont{A.~L.} \bibnamefont{Aquila}},
  \bibinfo{author}{\bibfnamefont{E.~M.} \bibnamefont{Gullikson}},
  \bibinfo{author}{\bibfnamefont{D.~T.} \bibnamefont{Attwood}},
  \bibinfo{author}{\bibfnamefont{R.}~\bibnamefont{Kienberger}},
  \bibnamefont{et~al.}, \bibinfo{journal}{Science}
  \textbf{\bibinfo{volume}{320}}, \bibinfo{pages}{1614} (\bibinfo{year}{2008}).

\bibitem[{\citenamefont{Sekikawa et~al.}(2004)\citenamefont{Sekikawa, Kosuge,
  Kanai, and Watanabe}}]{SEKIKAWA2004}
\bibinfo{author}{\bibfnamefont{T.}~\bibnamefont{Sekikawa}},
  \bibinfo{author}{\bibfnamefont{A.}~\bibnamefont{Kosuge}},
  \bibinfo{author}{\bibfnamefont{T.}~\bibnamefont{Kanai}}, \bibnamefont{and}
  \bibinfo{author}{\bibfnamefont{S.}~\bibnamefont{Watanabe}},
  \bibinfo{journal}{Nature} \textbf{\bibinfo{volume}{432}},
  \bibinfo{pages}{605} (\bibinfo{year}{2004}).

\bibitem[{\citenamefont{Corkum and Krausz}(2007)}]{CORKUM2007}
\bibinfo{author}{\bibfnamefont{P.~B.} \bibnamefont{Corkum}} \bibnamefont{and}
  \bibinfo{author}{\bibfnamefont{F.}~\bibnamefont{Krausz}},
  \bibinfo{journal}{Nat. Physics} \textbf{\bibinfo{volume}{3}},
  \bibinfo{pages}{381} (\bibinfo{year}{2007}).

\bibitem[{\citenamefont{Mauritsson et~al.}(2010)\citenamefont{Mauritsson,
  Remetter, Swoboda, Klunder, L'Huillier, Schafer, Ghafur, Kelkensberg, Siu,
  Johnsson et~al.}}]{MAURITSSON2010}
\bibinfo{author}{\bibfnamefont{J.}~\bibnamefont{Mauritsson}},
  \bibinfo{author}{\bibfnamefont{T.}~\bibnamefont{Remetter}},
  \bibinfo{author}{\bibfnamefont{M.}~\bibnamefont{Swoboda}},
  \bibinfo{author}{\bibfnamefont{K.}~\bibnamefont{Klunder}},
  \bibinfo{author}{\bibfnamefont{A.}~\bibnamefont{L'Huillier}},
  \bibinfo{author}{\bibfnamefont{K.~J.} \bibnamefont{Schafer}},
  \bibinfo{author}{\bibfnamefont{O.}~\bibnamefont{Ghafur}},
  \bibinfo{author}{\bibfnamefont{F.}~\bibnamefont{Kelkensberg}},
  \bibinfo{author}{\bibfnamefont{W.}~\bibnamefont{Siu}},
  \bibinfo{author}{\bibfnamefont{P.}~\bibnamefont{Johnsson}},
  \bibnamefont{et~al.}, \textbf{\bibinfo{volume}{arXiv:1001.1085v2}}
  (\bibinfo{year}{2010}).

\bibitem[{\citenamefont{Khundkar and Zewail}(1990)}]{KHUNDKAR1990}
\bibinfo{author}{\bibfnamefont{L.~R.} \bibnamefont{Khundkar}} \bibnamefont{and}
  \bibinfo{author}{\bibfnamefont{A.~H.} \bibnamefont{Zewail}},
  \bibinfo{journal}{Annu. Rev. Phys. Chem.} \textbf{\bibinfo{volume}{41}},
  \bibinfo{pages}{15} (\bibinfo{year}{1990}).

\bibitem[{\citenamefont{Scherer et~al.}(1991)\citenamefont{Scherer, Carlson,
  Matro, Du, Ruggiero, Romerorochin, Cina, Fleming, and Rice}}]{SCHERER1991}
\bibinfo{author}{\bibfnamefont{N.~F.} \bibnamefont{Scherer}},
  \bibinfo{author}{\bibfnamefont{R.~J.} \bibnamefont{Carlson}},
  \bibinfo{author}{\bibfnamefont{A.}~\bibnamefont{Matro}},
  \bibinfo{author}{\bibfnamefont{M.}~\bibnamefont{Du}},
  \bibinfo{author}{\bibfnamefont{A.~J.} \bibnamefont{Ruggiero}},
  \bibinfo{author}{\bibfnamefont{V.}~\bibnamefont{Romerorochin}},
  \bibinfo{author}{\bibfnamefont{J.~A.} \bibnamefont{Cina}},
  \bibinfo{author}{\bibfnamefont{G.~R.} \bibnamefont{Fleming}},
  \bibnamefont{and} \bibinfo{author}{\bibfnamefont{S.~A.} \bibnamefont{Rice}},
  \bibinfo{journal}{J. Chem. Phys.} \textbf{\bibinfo{volume}{95}},
  \bibinfo{pages}{1487} (\bibinfo{year}{1991}).

\bibitem[{\citenamefont{Feuerstein et~al.}(2007)\citenamefont{Feuerstein,
  Ergler, Rudenko, Zrost, Schroter, Moshammer, Ullrich, Niederhausen, and
  Thumm}}]{FEUERSTEIN2007}
\bibinfo{author}{\bibfnamefont{B.}~\bibnamefont{Feuerstein}},
  \bibinfo{author}{\bibfnamefont{T.}~\bibnamefont{Ergler}},
  \bibinfo{author}{\bibfnamefont{A.}~\bibnamefont{Rudenko}},
  \bibinfo{author}{\bibfnamefont{K.}~\bibnamefont{Zrost}},
  \bibinfo{author}{\bibfnamefont{C.~D.} \bibnamefont{Schroter}},
  \bibinfo{author}{\bibfnamefont{R.}~\bibnamefont{Moshammer}},
  \bibinfo{author}{\bibfnamefont{J.}~\bibnamefont{Ullrich}},
  \bibinfo{author}{\bibfnamefont{T.}~\bibnamefont{Niederhausen}},
  \bibnamefont{and} \bibinfo{author}{\bibfnamefont{U.}~\bibnamefont{Thumm}},
  \bibinfo{journal}{Phys. Rev. Lett.} \textbf{\bibinfo{volume}{99}},
  \bibinfo{pages}{153002} (\bibinfo{year}{2007}).

\bibitem[{\citenamefont{Winter et~al.}(2009)\citenamefont{Winter, Schmidt, and
  Thumm}}]{WINTER2009}
\bibinfo{author}{\bibfnamefont{M.}~\bibnamefont{Winter}},
  \bibinfo{author}{\bibfnamefont{R.}~\bibnamefont{Schmidt}}, \bibnamefont{and}
  \bibinfo{author}{\bibfnamefont{U.}~\bibnamefont{Thumm}},
  \bibinfo{journal}{Phys. Rev. A} \textbf{\bibinfo{volume}{80}},
  \bibinfo{pages}{031401} (\bibinfo{year}{2009}).

\bibitem[{\citenamefont{Jones et~al.}(1993)\citenamefont{Jones, Raman,
  Schumacher, and Bucksbaum}}]{JONES1993}
\bibinfo{author}{\bibfnamefont{R.~R.} \bibnamefont{Jones}},
  \bibinfo{author}{\bibfnamefont{C.~S.} \bibnamefont{Raman}},
  \bibinfo{author}{\bibfnamefont{D.~W.} \bibnamefont{Schumacher}},
  \bibnamefont{and} \bibinfo{author}{\bibfnamefont{P.~H.}
  \bibnamefont{Bucksbaum}}, \bibinfo{journal}{Phys. Rev. Lett.}
  \textbf{\bibinfo{volume}{71}}, \bibinfo{pages}{2575} (\bibinfo{year}{1993}).

\bibitem[{\citenamefont{Wollenhaupt et~al.}(2002)\citenamefont{Wollenhaupt,
  Assion, Liese, Sarpe-Tudoran, Baumert, Zamith, Bouchene, Girard, Flettner,
  Weichmann et~al.}}]{WOLLENHAUPT2002}
\bibinfo{author}{\bibfnamefont{M.}~\bibnamefont{Wollenhaupt}},
  \bibinfo{author}{\bibfnamefont{A.}~\bibnamefont{Assion}},
  \bibinfo{author}{\bibfnamefont{D.}~\bibnamefont{Liese}},
  \bibinfo{author}{\bibfnamefont{C.}~\bibnamefont{Sarpe-Tudoran}},
  \bibinfo{author}{\bibfnamefont{T.}~\bibnamefont{Baumert}},
  \bibinfo{author}{\bibfnamefont{S.}~\bibnamefont{Zamith}},
  \bibinfo{author}{\bibfnamefont{M.~A.} \bibnamefont{Bouchene}},
  \bibinfo{author}{\bibfnamefont{B.}~\bibnamefont{Girard}},
  \bibinfo{author}{\bibfnamefont{A.}~\bibnamefont{Flettner}},
  \bibinfo{author}{\bibfnamefont{U.}~\bibnamefont{Weichmann}},
  \bibnamefont{et~al.}, \bibinfo{journal}{Phys. Rev. Lett.}
  \textbf{\bibinfo{volume}{89}}, \bibinfo{pages}{173001}
  (\bibinfo{year}{2002}).

\bibitem[{\citenamefont{Palacios et~al.}(2009)\citenamefont{Palacios, Rescigno,
  and McCurdy}}]{PALACIOS2009}
\bibinfo{author}{\bibfnamefont{A.}~\bibnamefont{Palacios}},
  \bibinfo{author}{\bibfnamefont{T.~N.} \bibnamefont{Rescigno}},
  \bibnamefont{and} \bibinfo{author}{\bibfnamefont{C.~W.}
  \bibnamefont{McCurdy}}, \bibinfo{journal}{Phys. Rev. Lett.}
  \textbf{\bibinfo{volume}{103}}, \bibinfo{pages}{253001}
  (\bibinfo{year}{2009}).

\bibitem[{\citenamefont{Nubbemeyer et~al.}(2008)\citenamefont{Nubbemeyer,
  Gorling, Saenz, Eichmann, and Sandner}}]{NUBBEMEYER2008}
\bibinfo{author}{\bibfnamefont{T.}~\bibnamefont{Nubbemeyer}},
  \bibinfo{author}{\bibfnamefont{K.}~\bibnamefont{Gorling}},
  \bibinfo{author}{\bibfnamefont{A.}~\bibnamefont{Saenz}},
  \bibinfo{author}{\bibfnamefont{U.}~\bibnamefont{Eichmann}}, \bibnamefont{and}
  \bibinfo{author}{\bibfnamefont{W.}~\bibnamefont{Sandner}},
  \bibinfo{journal}{Phys. Rev. Lett.} \textbf{\bibinfo{volume}{101}},
  \bibinfo{pages}{233001} (\bibinfo{year}{2008}).

\bibitem[{\citenamefont{Eichmann et~al.}(2009)\citenamefont{Eichmann,
  Nubbemeyer, Rottke, and Sandner}}]{EICHMANN2009}
\bibinfo{author}{\bibfnamefont{U.}~\bibnamefont{Eichmann}},
  \bibinfo{author}{\bibfnamefont{T.}~\bibnamefont{Nubbemeyer}},
  \bibinfo{author}{\bibfnamefont{H.}~\bibnamefont{Rottke}}, \bibnamefont{and}
  \bibinfo{author}{\bibfnamefont{W.}~\bibnamefont{Sandner}},
  \bibinfo{journal}{Nature} \textbf{\bibinfo{volume}{461}},
  \bibinfo{pages}{1261} (\bibinfo{year}{2009}).

\bibitem[{\citenamefont{Javanainen et~al.}(1988)\citenamefont{Javanainen,
  Eberly, and Su}}]{JAVANAINEN1988}
\bibinfo{author}{\bibfnamefont{J.}~\bibnamefont{Javanainen}},
  \bibinfo{author}{\bibfnamefont{J.~H.} \bibnamefont{Eberly}},
  \bibnamefont{and} \bibinfo{author}{\bibfnamefont{Q.~C.} \bibnamefont{Su}},
  \bibinfo{journal}{Phys. Rev. A} \textbf{\bibinfo{volume}{38}},
  \bibinfo{pages}{3430} (\bibinfo{year}{1988}).

\bibitem[{\citenamefont{Fleck et~al.}(1976)\citenamefont{Fleck, Morris, and
  Feit}}]{FLECK1976}
\bibinfo{author}{\bibfnamefont{J.~A.} \bibnamefont{Fleck}},
  \bibinfo{author}{\bibfnamefont{J.~R.} \bibnamefont{Morris}},
  \bibnamefont{and} \bibinfo{author}{\bibfnamefont{M.~D.} \bibnamefont{Feit}},
  \bibinfo{journal}{Appl. Phys.} \textbf{\bibinfo{volume}{10}},
  \bibinfo{pages}{129} (\bibinfo{year}{1976}).

\bibitem[{\citenamefont{Feit et~al.}(1982)\citenamefont{Feit, Fleck, and
  Steiger}}]{FEIT1982}
\bibinfo{author}{\bibfnamefont{M.~D.} \bibnamefont{Feit}},
  \bibinfo{author}{\bibfnamefont{J.~A.} \bibnamefont{Fleck}}, \bibnamefont{and}
  \bibinfo{author}{\bibfnamefont{A.}~\bibnamefont{Steiger}},
  \bibinfo{journal}{J. Comput. Phys.} \textbf{\bibinfo{volume}{47}},
  \bibinfo{pages}{412} (\bibinfo{year}{1982}).

\bibitem[{\citenamefont{Pfeifer et~al.}(2004)\citenamefont{Pfeifer, Walter,
  Gerber, Emelin, Ryabikin, Chernobrovtseva, and Sergeev}}]{PFEIFER2004A}
\bibinfo{author}{\bibfnamefont{T.}~\bibnamefont{Pfeifer}},
  \bibinfo{author}{\bibfnamefont{D.}~\bibnamefont{Walter}},
  \bibinfo{author}{\bibfnamefont{G.}~\bibnamefont{Gerber}},
  \bibinfo{author}{\bibfnamefont{M.~Y.} \bibnamefont{Emelin}},
  \bibinfo{author}{\bibfnamefont{M.~Y.} \bibnamefont{Ryabikin}},
  \bibinfo{author}{\bibfnamefont{M.~D.} \bibnamefont{Chernobrovtseva}},
  \bibnamefont{and} \bibinfo{author}{\bibfnamefont{A.~M.}
  \bibnamefont{Sergeev}}, \bibinfo{journal}{Phys. Rev. A}
  \textbf{\bibinfo{volume}{70}}, \bibinfo{pages}{013805}
  (\bibinfo{year}{2004}).

\end{thebibliography}

\end{document}